\documentclass[twocolumn]{htl-author}
\usepackage[T1]{fontenc}
\usepackage{multirow}
\usepackage{booktabs}
\usepackage{tabularx}
\usepackage{amsmath}
\usepackage{caption}
\usepackage{cleveref}
\usepackage{stfloats}
\usepackage{float}

\crefname{figure}{Fig.}{Figs.}
\usepackage{graphicx}
\newcolumntype{Y}{>{\centering\arraybackslash}X}

\usepackage{xcolor}

\definecolor{rebuttal}{RGB}{0, 0, 0} % This defines an orange color

\begin{document}

\title{RGB to Hyperspectral: Spectral Reconstruction for Enhanced Surgical Imaging}

\author{Tobias Czempiel$^{1,2,3}$, Alfie Roddan$^{2}$, Maria Leiloglou$^{1,2,3}$, Zepeng Hu$^{2}$, Kevin O'Neill$^{4}$, Giulio Anichini$^{2,4}$, Danail Stoyanov$^{3}$, Daniel Elson$^{2}$}

\address{
E-mail: t.czempiel@ic.ac.uk \\
$^{1}$EnAcuity Limited, London UK\\
$^{2}$Hamlyn Centre for Robotic Surgery, Department of Surgery and Cancer, Imperial College London\\
$^{3}$Wellcome/EPSRC Centre for Interventional and Surgical Sciences (WEISS) and Department of Computer Science, University College London\\
$^{4}$Department of Surgery and Cancer, Healthcare NHS Trust, Imperial College London}

\historydate{Published in Healthcare Technology Letters; Received on xxx; Revised on xxx}

\abstract{This study investigates the reconstruction of hyperspectral signatures from RGB data to enhance surgical imaging, utilizing the publicly available HeiPorSPECTRAL dataset from porcine surgery and an in-house neurosurgery dataset. Various architectures based on convolutional neural networks (CNNs) and transformer models are evaluated using comprehensive metrics. Transformer models exhibit superior performance in terms of RMSE, SAM, PSNR and SSIM by effectively integrating spatial information to predict accurate spectral profiles, encompassing both visible and extended spectral ranges. Qualitative assessments demonstrate the capability to predict spectral profiles critical for informed surgical decision-making during procedures. Challenges associated with capturing both the visible and extended hyperspectral ranges are highlighted using the MAE, emphasizing the complexities involved. The findings open up the new research direction of hyperspectral reconstruction for surgical applications and clinical use cases in real-time surgical environments.}

\maketitle

\section{Introduction}
\begin{figure}[h!]
    \centering
    \includegraphics[width=8cm]{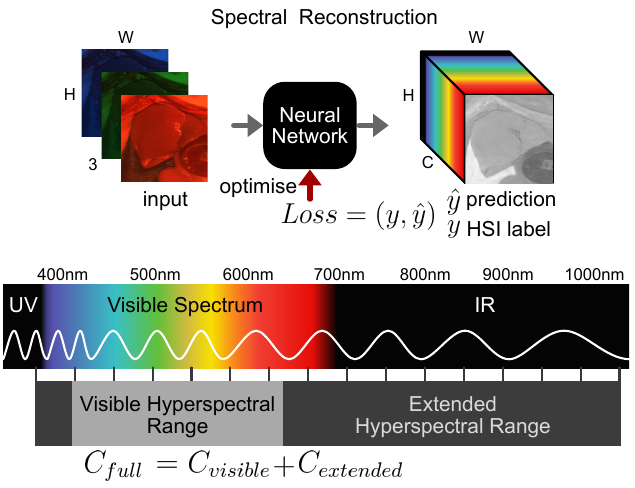}
    \captionsetup{width=\linewidth}
    \caption{At the top the prediction step from three channel (C) RGB to HSI $\hat{y}$ is shown. The Neural Network is optimised using the Loss between $\hat{y}$ and the ground truth HSI data $y$. At the bottom of the figure the different spectral ranges ($C_{full}, C_{visible}, C_{extended}$) are visualized that are used to for evaluation.}
    \label{fig:overview}
\end{figure}

Hyperspectral imaging (HSI) is increasingly recognized in surgical applications for its ability to reveal intricate details that conventional imaging methods often overlook \cite{Lu2014, ANICHINI2024108293}. Unlike traditional cameras, which capture a broad range of wavelengths within the Red, Green, and Blue (RGB) parts of the electromagnetic spectrum, HSI extends this capability by detecting narrower wavelength bands over a larger range of the electromagnetic spectrum \cite{vercauteren, Chang2007}. The resultant hyperspectral image contains more than three channels, each representing one narrow spectral band of reflected light. In \cref{fig:overview} the electomagnetic spectrum is visualized with a corresponding HSI image with C the number of wavelengths in the channel dimension. Depending on the hardware C can vary with different mininmum and maximum wavelength. This capability allows HSI to unveil critical information, such as the presence of tumors and detailed functional insights like blood hemodynamics, without the need for contrast agents like indocyanine green (ICG) \cite{CLANCY2020101699}.

Deep learning (DL) is often used as a tool to analyse this data and allow for predictions of multiple downstream tasks such as classification, localisation, regression or segmentation of biological tissue \cite{Cui2022}. Despite its potential, the widespread adoption of HSI in surgery is hindered by significant challenges associated with specialized HSI cameras. These cameras are costly, bulky, and often compromise on factors such as spatial resolution, spectral resolution, frame rate, and overall size and cost \cite{Cui2022}.
To algorithmically alleviate some of these limitations, the vision community has proposed various concepts. One approach involves reconstructing spatial information from low-resolution RGB images to high-resolution RGB images. By capturing the semantic content of low-resolution images, it is possible to reconstruct high-resolution images with much better performance than simple upscaling could provide.
Motivated by this success, researchers have explored the reconstruction of not only spatial information but also spectral information. RGB can be seen as a low-resolution, channel-wise representation, and the goal of the reconstruction task is to recover the much higher spectral resolution of multispectral (greater than three channels) and hyperspectral images (typically greater than 20 channels). In \cref{fig:overview} the concept is visualized where a Neural Network is used to reconstruct HSI from the RGB information.

An emerging approach from the vision community has gained attention: reconstructing hyperspectral information from RGB images \cite{Zhang2022}. This approach, pioneered in challenges like the NTIRE challenge \cite{Arad2022} utilizes DL models based on CNN or more recently transformer architectures, enabling existing RGB imaging systems to provide more comprehensive spectral data while circumventing the drawbacks of specialized HSI hardware. By exploring these contributions, this paper aims to advance the understanding and application of RGB to HSI reconstruction in surgical contexts. 

This paper presents a pioneering analysis in the field of surgical HSI reconstruction, focusing on several key contributions. First, we introduce a comparative architectural framework for RGB to HSI reconstruction methods, thoroughly evaluating a range of architectures \textcolor{rebuttal}{simple fully connected layers with non-linearities such as ReLU to advanced deep learning architectures like CNNs and transformers.} Second, we conduct an in-depth metric analysis to assess the quality and fidelity of spectral reconstruction achieved through RGB imaging compared to true hyperspectral data. This includes a detailed examination of reconstruction performance across different spectral ranges, highlighting the strengths and limitations of each model. Finally, we provide qualitative insights into the spatial reconstruction capabilities of these models on two distinct datasets — one focused on neurosurgical applications and the other on colorectal procedures. This comprehensive evaluation assesses the quality and effectiveness of the reconstructed hyperspectral images across various wavelengths, highlighting the models’ performance across different medical specialties and procedural contexts. These contributions aim to bridge the gap between advancements in the vision community and practical medical imaging applications, paving the way for cost-effective and detailed spectral information acquisition in surgical environments.

\section{Related Work}
% Flow of related work:

% what can HSI offer
% what we want to do with HSI (CAI)
    % detection, statistical models - skin classification
    % segmentation, machine learning models 
    % functional 

%  There exist method to go directly from RGB to functional 
% how we get hsi from RGB
    % there exists a challenge Ntire
    % CNN model
    % Transformers 
% available datasets for surgery  
    % HIB dataset
    % Dan dataset
    % HeiPorSPECTRAL
% summary

% CLINICAL MOTIVATION WHY HSI IS IMPORTANT
Surgical HSI offers the ability to capture light information not only in greater detail within the visible spectrum but also beyond it. This detailed spectral information facilitates the separation of biological tissues, enhancing the predictive capabilities of statistical and DL models for functional information \cite{Lu2014}. HSI has been used for many different clinical applications ranging from advanced dermatology \cite{Hetz2024AdvancingDD} and wound care \cite{Saiko2020-ra} to intraoperative tumour margin delineation \cite{Akbari2011}, retinal vasculature segmentation \cite{garifullin2018hyperspectral} and determination of the transection margin during colorectal resection \cite{JansenWinkeln2019}. Additionally, HSI can be used to predict functional information, this includes specific, actionable insights such as tissue oxygenation levels or blood flow dynamics \cite{Holmer2018, Kohler2021} to be used by the surgeon. Specifically, surgeons may wish to visualize or estimate Tissue oxygenation (StO2) or other relevant functional informations such as Tissue Hemoglobin Index (THI) or Near Infrared Perfusion Index (NPI) \cite{Dietrich2021} in order to aid decision making during procedures. While some work has attempted to derive this functional information from RGB images \cite{Jones2017}, the approach has not been widely explored and tends to lack ground truth labels, leading to non-robust models across different settings. Nevertheless, the success of these initial efforts serves as a strong motivation for our research.

In summary, the clinical applications of HSI are extensive, making the reconstruction of HSI information from RGB highly beneficial. Given the higher information density of HSI images compared to standard three-channel RGB images, this reconstruction is an ill-posed problem. In the vision domain, the reconstruction of HSI from RGB has been explored using various algorithm types. Statistical methods such as regression and sparse coding \cite{Wu2017InDO}, as well as advanced CNN approaches \cite{Galliani2017LearnedSS,Stiebel2018}, have been investigated. Recently, a challenge (NTIRE \cite{Arad2022}) compared different deep learning-based approaches, utilizing a variety of architecture types and loss functions. With the emergence of transformers, new methods for reconstructing hyperspectral information have also been developed \cite{Cai2022}, showcasing the potential of these advanced models to enhance spectral reconstruction performance.

The development of a RGB to HSI reconstruction model for surgical applications depends heavily on the availability of data. The dataset must not only contain HSI but also RGB information that have been spatially registered together or reconstructed. Spectralpaca \cite{Ayala2024}, a recently introduced hyperspectral video dataset, is designed for human perfusion monitoring. It features spectral video recordings of ten healthy participants in different physiological states. While this dataset addresses medical applications, it does not specifically pertain to surgical contexts. The hyperspectral imaging benchmark (HIB) \cite{Leon2023} dataset is comprised of hyperspectral and RGB images during neurosurgery, offering comprehensive insights into tissue composition and pathology. Additionally, the HeiPorSPECTRAL dataset \cite{Studier-Fischer2023} contributes valuable spectral data of porcine organs offering a much larger dataset and has been used to develop advancde deep learning based methods (details in \cref{sec:HeiPorSPECTRAL}).

However, despite the critical importance of in vivo validation for clinical translation of HSI, the biomedical field faces a significant challenge due to the scarcity of publicly available datasets. Nonetheless, these datasets provide a solid groundwork for advancing surgical interventions, allowing surgeons to harness comprehensive spectral information to improve patient care and surgical outcomes. Our goal is to integrate the advancements of HSI for surgery with methodological innovations in RGB to hyperspectral reconstruction. In the following sections, we detail our approach to achieve this goal.

%%% other

% Unlike RGB imaging, HSI allows for tissue classification \cite{Studier-Fischer2022} on a per-pixel basis due to the rich spectral information available using statistical models or DL-based approaches \cite{Lu2014}. 

% Dual-modality endoscopic probe for tissue surface shape reconstruction and hyperspectral imaging enabled by deep neural networks \cite{Lin2018}

% % FUNCTIONAL ANALYSIS USING HSI FOR MEDICAL
% Hyperspectral reconstruction from RGB images for vein visualization \cite{Sharma2020}

% Hyper-Skin: A Hyperspectral Dataset for Reconstructing Facial Skin-Spectra from RGB Images \cite{Ng2023}

\section{Method}
The purpose of this study is to investigate whether RGB to HSI reconstruction is viable for surgical HSI. To achieve this, we compare various DL architectures from both the imaging domain and the spectral reconstruction domain.

\subsection{Architecture}
For the architecture choice we wanted to find a mix between advanced DL models based on CNNs and Transformers but also simpler models with fewer parameters and non-linearities to investigate the necessity and improvements between different architecture complexities.

\subsubsection{PixelFeatureNet}
The PixelFeatureNet employs a simple per-pixel design analysing the spectra across the entire range without considering any spacial neighborhood relationships. It consists of four fully-connected layers where the channel dimensionality per pixel increases progressively. The input comprises 3 channels (RGB), and the output dimensionality is C, representing the number of channels for the dataset. ReLU activation functions are used throughout.

\subsubsection{LocalFeatureNet}
The LocalFeatureNet replaces the fully-connected layers of PixelFeatureNet with convolutional layers, each utilizing a fixed kernel size of 3 across all four layers. This modification introduces local neighborhood information, aimed at enhancing spectral reconstruction capabilities while maintaining a lightweight model design.

\subsubsection{UNET} 
Additionally, we incorporated a standard U-Net architecture \cite{Ronneberger2015}, renowned for its effectiveness in medical segmentation and reconstruction tasks. The U-Net’s design, featuring bottleneck layers and skip connections, facilitates capturing both global context and detailed local features, thereby aiding in precise spectral reconstruction. We use a ResNet50 \cite{he2015deepresiduallearningimage} backbone pretrained on Imagenet \cite{deng2009imagenet}. 
% Based on resnet50 and imagenet weights. 

\subsubsection{MST++}
For our final model, we selected a transformer-based architecture \cite{Vaswani2017} called MST++ \cite{Cai2022}. This model has demonstrated significant potential in reconstructing real-world images and stands as the state-of-the-art on the NTIRE dataset. The transformer architecture of MST++ was designed to leverage global dependencies effectively, by utilising spectral-wise self-attention it specifically targets the HSI modality, potentially improving spectral reconstruction accuracy.

Overall the selection of models allows us to explore different architectural paradigms and their impact on hyperspectral image reconstruction tasks.

\subsection{Metrics}
% In the reconstruction of RGB images, several metrics are used to evaluate the quality and accuracy of the reconstructed images. 

There are several metrics to evaluate the quality of the reconstruction of hyperspectral images. Hyperspectral data has the unique characteristic of being both an image and also a composition of spectra. For this reason we can not only evaluate image level values with metrics like mean absolute error (MAE), root mean squared error (RMSE), peak signal-to-noise ratio (PSNR) and structural similarity index (SSIM) but also compare spectral quality with spectral angle mapping (SAM). The image level metrics aim to evaluate the reconstructed images' absolute values and texture whereas SAM focuses on similarity at a spectral level.

\textcolor{rebuttal}{
For an RGB image $X$ with a ground truth hyperspectral image $Y$, each image has a total of $N = H \times W \times C$ pixels, where $H$ is the height, $W$ is the width, and $C$ is the number of channels. A reconstructed image $\hat{y}$ also contains $N$ pixels, with each pixel $\hat{y}_i$ corresponding to a ground truth observation $y_i$. Each pixel in $N$ contributes equally to the metric, regardless of its channel, representing a micro-average approach. We further report channel-wise metrics in \cref{sec:Reconstruction for different hyperspectral ranges} to provide a more detailed analysis.}

\subsubsection{Mean Absolute Error (MAE)}
The MAE is calculated as and is also known as the L1 Norm:
\[ \text{MAE}(y, \hat{y}) = \frac{1}{N} \sum_{i=1}^{n} |y_i - \hat{y}_i| \]

\subsubsection{Root Mean Square Error (RMSE)}
RMSE is calculated as the square root of the average of the squared differences between the predicted and observed values. In hyperspectral reconstruction, RMSE quantifies the error between the reconstructed spectral data and the ground truth spectral data for each pixel.

\[
\text{RMSE}(y, \hat{y}) = \sqrt{\frac{1}{N} \sum_{i=1}^{N} (y_i - \hat{y}_i)^2}
\]

\subsubsection{Peak Signal-to-Noise Ratio (PSNR)}
Peak Signal-to-Noise Ratio (PSNR) is a measure of the quality of the reconstructed image compared to the original image. It is expressed in decibels (dB) and is inversely proportional to the mean squared error. Higher PSNR values indicate better reconstruction quality.

\[
\text{PSNR} = 10 \cdot \log_{10} \left( \frac{\text{MAX}^2}{\text{MSE}} \right)
\]

where \(\text{MAX}\) is the maximum possible pixel value of the image, and \(\text{MSE}\) is the mean squared error.

\subsubsection{Spectral Angle Mapper (SAM)}
Spectral Angle Mapper (SAM) is a spectral classification method that measures the spectral similarity between two spectra by calculating the angle between them. It is used to compare the similarity of the spectral signatures of the reconstructed and ground truth images. Smaller angles indicate higher similarity.

\[
\text{SAM}(y, \hat{y}) = \cos^{-1} \left( \frac{\sum_{i=1}^{N} y_i \hat{y}_i}{\sqrt{\sum_{i=1}^{N} y_i^2} \cdot \sqrt{\sum_{i=1}^{N} \hat{y_i}^2}} \right)
\]

\subsubsection{Structural Similarity Index (SSIM)}
Structural Similarity Index (SSIM) is a perceptual metric that measures the similarity between two images. It considers changes in structural information, luminance, and contrast. SSIM values range from -1 to 1, with 1 indicating perfect similarity.
\[
\text{SSIM}(y, \hat{y}) = \frac{(2\mu_y \mu_{\hat{y}} + C_1)(2\sigma_{y\hat{y}} + C_2)}{(\mu_y^2 + \mu_{\hat{y}}^2 + C_1)(\sigma_y^2 + \sigma_{\hat{y}}^2 + C_2)}
\]
where \(\mu_y\) and \(\mu_{\hat{y}}\) are the average pixel values of images \(y\) and prediction \(\hat{y}\), \(\sigma_y^2\) and \(\sigma_{\hat{y}}^2\) are the variances, \(\sigma_{y\hat{y}}\) is the covariance, and \(C_1\) and \(C_2\) are constants to stabilize the division.

These metrics collectively provide a comprehensive evaluation of the quality of hyperspectral reconstruction, assessing both pixel-wise accuracy and structural fidelity.

\subsection{Analysis of Reconstruction in Different Wavelength Ranges}

The reconstruction of hyperspectral information is a severely ill-posed problem \cite{Zhang2022}. However, the difficulty of reconstructing the channels C varies depending on the wavelengths  $\lambda$  within  C. We hypothesise that hyperspectral wavelengths within the visible range are easier to reconstruct as RGB images contain a lot of macro information of these wavelengths. Conversely, wavelengths in the ultraviolet (UV) or infrared (IR) ranges are more challenging to reconstruct due to the lack of spectral information in these regions in the input RGB images. We consider the full hyperspectral range from the lowest wavelength in the hyperspectral dataset $\lambda_{min}$ to the highest wavelength $\lambda_{max}$ of a dataset as follows:

\[
\lambda \in [\lambda_{min},\lambda_{max}] \quad (C_{\text{full}})
\]

\subsubsection{Visible Hyperspectral Range}  
The visible hyperspectral range is defined between 
\[
400 \, \text{nm} \leq \lambda \leq 680 \, \text{nm} \quad (C_{\text{visible}})
\]
This aligns with the common RGB sensor ranges \cite{Arad2022}. We expect the reconstruction of these wavelength channels to be easier because it can be thought of as interpolation between the channels of RGB. While still ill-posed, it is not as severely ill-posed.

\subsubsection{Extended Hyperspectral Range}  
As visualized in \cref{fig:overview}, wavelengths below the blue spectrum fall into the ultraviolet light range (\(\lambda < 460 \, \text{nm}\)), and those above the visible red spectrum fall into the infrared light range (\(\lambda > 680 \, \text{nm}\)). There is limited information about these ranges present in the RGB images used as an information source for reconstruction. Therefore, we define the extended hyperspectral range as follows:

\[
\lambda < 400 \, \text{nm} \quad \text{or} \quad \lambda > 680 \, \text{nm} \quad (C_{\text{extended}})
\]

This categorization highlights the challenges faced in reconstructing hyperspectral data in different wavelength ranges and sets the stage for further analysis of reconstruction methods.

\section{Experimental Setup}
\subsection{Datasets}
For our study, we selected two distinct datasets representing different surgical interventions and data acquisition hardware.
We aim to leverage diverse surgical scenarios and varied hardware setups to explore the robustness and applicability of our methods. The datasets chosen offer complementary insights into surgical imaging and enable comprehensive evaluations of our approach. \textcolor{rebuttal}{We conducted an investigation into the results obtained by training models on the datasets individually without applying any transfer learning, as shown in \cref{tab:methods_comparison}. Additionally, we explored the impact of pre-training on one dataset followed by fine-tuning on the other, which is detailed in \cref{tab:transfer_learning}.}

\subsubsection{MSI Brain}
The MSI Brain dataset is comprised of in-house neurosurgery data captured using a liquid crystal tunable filter system integrated with a Zeiss operating microscope, illuminated by a standard xenon surgical light source. The RGB data, with a resolution of 1920x1080 pixels, covers a comprehensive area of the surgical field. Following registration to hyperspectral imaging (HSI), the RGB data is masked to match the spatial dimensions of the HSI data. The HSI data itself spans a spectral range from 460 nm to 720 nm with a 10 nm sampling resolution, captured by a monochromatic camera with a spatial resolution of 576x768 pixels. Due to the multispectral acquisition lasting several seconds, motion artifacts were encountered, necessitating coregistration of the hyperspectral cubes. This preprocessing step involved initial SIFT registration followed by more precise RANSAC-Flow alignment to ensure high spatial registration accuracy. The dataset of 225 hyperspectral images was randomly divided into training (135 images, 60\%), validation (22 images, 10\%), and test sets (68 images, 30\%). As part of preprocessing, a histogram equalization step was applied to enhance contrast and standardize image quality, aiming to improve visualization and facilitate subsequent image analysis tasks in neurosurgery

% RGB Range 460nm to 680
% Extended Range: 690 to 720

\subsubsection{HeiPorSPECTAL}\label{sec:HeiPorSPECTRAL}
The HeiPorSPECTAL dataset \cite{Studier-Fischer2023} comprises HSI data from 20 pigs, acquired at Heidelberg University Hospital. This dataset includes image annotations for 20 different organs, providing a valuable resource for medical imaging research. The acquisition system uses the Tivita® Tissue HSI camera system (Diaspective Vision GmbH, Am Salzhaff, Germany), capturing a spectral range from 500 nm to 1000 nm without overlaps. The resulting image resolution of each hypercube is 640x480x100 pixel. Corresponding RGB images are reconstructed from the HSI data by aggregating spectral channels that capture red, green, and blue light \cite{Holmer2018}. As the RGB is reconstructed from the hyperspectral data no registration is necessary between RGB and HSI as they are already aligned. 
The HSI cubes were further preprocessed with L1 normalization. L1 normalization scales the data so that the sum of the absolute reflectance values for each pixel equals 1, ensuring that each reflectance spectrum is proportionally scaled. We use the official split provided by the authors (\textit{Split2}) with 827 images for training, 1260 for validation and 1366 for testing. 

\subsection{Training details}
We trained all our models for 100 epochs with a learning rate of \(1 \times 10^{-4}\) and chose the best performing model based on the validation set and evaluate the model using the unseen test set. We utilized the Adam optimizer with betas set to \( (0.9, 0.999) \), and employed a cosine annealing learning rate scheduler initialized with \( \text{total\_iteration} \) and \( \eta_{\text{min}} = 1 \times 10^{-6} \) for training. To enhance data variability and mitigate overfitting to spatial cues, we employed standard geometric augmentation techniques, including resizing to 288 by 480 pixels in the spatial dimension, and applying random flipping, shifting, scaling, and rotation during training.

For the loss functions, we used the L1 loss (equivalent to Mean Absolute Error) and the Mean Relative Absolute Error (MRAE) frequently used for spectral reconstruction \cite{Cai2022}, defined as:
\[
Loss_{MRAE}(y, \hat{y}) = \frac{1}{N} \sum_{i=1}^{N} \left| \frac{y_i - \hat{y}_i}{y_i} \right|
\]
where \( y_i \) and \( \hat{y}_i \) represent the ground truth and predicted values, respectively, for \( N \) samples. The division through the error with the label helps to normalize the error relative to the magnitude of the ground truth values, providing a measure of relative error that is independent of the scale of the data. 

Due to many instances in the MSI Brain dataset where $y_i$ equaled zero, the MRAE loss could not be used effectively. Therefore, we opted for the standard L1 loss function for the MSI Brain dataset.

\section{Results}

\begin{table*}[h!]
    \centering
    \caption{Comparison of Methods on Different Datasets using RMSE$\downarrow$, SAM$\downarrow$, SSIM$\uparrow$, and PSNR$\uparrow$. $\downarrow$ indicates that lower values are better and $\uparrow$ indicates that higher values are better \textcolor{rebuttal}{with $\mu$ the mean value and $\sigma$ the standard deviation}. Bold indicates best \textcolor{rebuttal}{mean ($\mu$)} value for the dataset. Additionally the number of learnable parameters is shown which is different for MSI Brain and HeiPorSPECTAL dataset due to the increased number of hyperspectral channels in the HeiPorSPECTAL dataset}
    \begin{tabularx}{0.7\textwidth}{>{\raggedright\arraybackslash}X >{\centering\arraybackslash}Y >{\centering\arraybackslash}Y >{\centering\arraybackslash}Y >{\centering\arraybackslash}Y >{\centering\arraybackslash}Y >{\centering\arraybackslash}Y} 
        \toprule
        \textbf{Architecture} & \textbf{\# Params} &\textbf{Metric} & \multicolumn{2}{c}{\textbf{MSI Brain}} & \multicolumn{2}{c}{\textbf{HeiPorSPECTAL}} \\
         & & & $\mu$ & $\sigma$ & $\mu$ & $\sigma$ \\
        \midrule
        \multirow{4}{*}{PixelFeatureNet} & \multirow{4}{*}{1.6K/4K} & RMSE & 0.0415 & $\pm 0.0143$ & 0.0026 & $\pm 0.0005$ \\ 
        & & SAM & 0.1985 & $\pm 0.0324$ & 0.1703 &$\pm 0.0574 $  \\
        & & PSNR & 28.1059 & $\pm 2.7969$ & 52.2938 & $\pm 2.6857$ \\
        & & SSIM & 0.7641 & $\pm 0.0303 $ & 0.9841 &$\pm 0.0087$ \\        
        \midrule
        \multirow{4}{*}{LocalFeatureNet} & \multirow{4}{*}{6.9K/9.3K} & RMSE & 0.0336 & $\pm 0.0135$ & 0.0025 & $\pm 0.0010$\\
         & & SAM & 0.1620 & $\pm 0.0376$ & 0.1578 & $\pm 0.0622$ \\
         & & PSNR & 30.0913 & $\pm 3.2425$ & 52.7109 & $\pm 2.9627 $ \\
         & & SSIM & 0.9484 & $\pm 0.0182$ & 0.9859 & $\pm 0.0092 $  \\
        \midrule
        \multirow{4}{*}{UNet} & \multirow{4}{*}{32.5M/32.5M} & RMSE & 0.0376 & $\pm 0.0133$ & 0.0029 & $\pm 0.0026 $ \\
        & & SAM & 0.1861 & $\pm 0.0540$ & \textbf{0.1258} & $\pm 0.0441$\\
        & & PSNR & 29.0298 & $\pm 3.0120$ & 52.5472 & $\pm 5.1437$ \\
        & & SSIM & 0.9334 & $\pm 0.0198$& 0.9498 & $\pm 0.0056$ \\

        \midrule
        \multirow{4}{*}{MST++} & \multirow{4}{*}{416K/5.6M} & RMSE & \textbf{0.0312} & $\pm 0.0102$ & \textbf{0.0021} & $\pm 0.0008$ \\
        & & SAM & \textbf{0.1528} & $\pm 0.0317$ & 0.1425 & $\pm 0.0316$ \\
        & & PSNR & \textbf{30.5883} & $\pm 2.8428$& \textbf{53.9625} & $\pm 2.8197$ \\
        & & SSIM & \textbf{0.9507} & $\pm 0.0145$& \textbf{0.9924} &  $\pm 0.0037$ \\
        % \bottomrule
    \end{tabularx}
    \label{tab:methods_comparison}
\end{table*}

% \begin{table*}[h!]
%     \centering
%     \caption{Comparison of Oxyplore on Different Datasets using RMSE$\downarrow$, SAM$\downarrow$, SSIM$\uparrow$, and PSNR$\uparrow$. $\downarrow$ indicates that lower values are better and $\uparrow$ indicates that higher values are better with $\mu$ the mean value and $\sigma$ the standard deviation.}
%     \begin{tabularx}{\textwidth}{>{\raggedright\arraybackslash}b{3cm} >{\centering\arraybackslash}b{3.0cm} >{\centering\arraybackslash}Y >{\centering\arraybackslash}Y >{\centering\arraybackslash}Y >{\centering\arraybackslash}Y} 
%         \toprule
%         \textbf{Architecture} &\textbf{Metric} & \multicolumn{2}{c}{\textbf{in-human neurosurgery}} & \multicolumn{2}{c}{\textbf{porcine visceral data}} \\
%           & & $\mu$ & $\sigma$ & $\mu$ & $\sigma$ \\       
%         \midrule
%         \multirow{4}{*}{\shortstack{\textbf{Oxyplore} \\ SpectralReconstruction}} & RootMeanSquareError & 0.0312 & $\pm 0.0102$ & 0.0021 & $\pm 0.0008$ \\
%         &  SpectralAngleMapper & 0.1528 & $\pm 0.0317$ & 0.1425 & $\pm 0.0316$ \\
%         &  PeakSignalToNoise & 30.5883 & $\pm 2.8428$& 53.9625 & $\pm 2.8197$ \\
%         &  StructuralSimilarityIndex & 0.9507 & $\pm 0.0145$& 0.9924 &  $\pm 0.0037$ \\
%         % \bottomrule
%     \end{tabularx}
%     \label{tab:methods_comparison}
% \end{table*}

In this section, we present our findings on RGB to hyperspectral reconstruction, both quantitatively across three distinct hyperspectral ranges and qualitatively with examples from both datasets.

\subsection{RGB to Hyperspectral Reconstruction Quantitative}
As depicted in \cref{tab:methods_comparison}, all models were successfully trained and fall within an acceptable range.

Notably, the straightforward PixelFeatureNet, comprising only four linear layers, performs competitively against the LocalFeatureNet, which incorporates spatial information. On the HeiPorSPECTRAL dataset, the RMSE results show a marginal difference (delta = 0.0001) between them. Both PixelFeatureNet and LocalFeatureNet demonstrate superior performance over UNet on the HeiPorSPECTRAL dataset by 0.0003 and 0.0004, respectively. These findings suggest that elaborate models with numerous parameters do not significantly enhance spectral reconstruction. Simple models, despite their lack of spatial modeling, achieve satisfactory results.

Interestingly, UNet exhibits the best overall performance on the HeiPorSPECTRAL dataset with an SAM of 0.1258, indicating that higher-level information and bottleneck features may aid in spectral consistency. Furthermore, the transformer-based MST++ model surpasses all others in terms of RMSE, PSNR, and SSIM, underscoring its efficacy in medical hyperspectral imaging. This suggests promising avenues for applying computer vision techniques from general domains like ImageNet to medical applications.

Conversely, on the MSI Brain dataset, PixelFeatureNet performs less favorably compared to other models, achieving the worst results across all metrics with an RMSE of 0.0415, compared to 0.0336 for the second-worst model. UNet, while ranking third, exhibits significantly lower SSIM (~17 difference) compared to the top-performing models. LocalFeatureNet surpasses UNet across all metrics, indicating that model complexity can potentially lead to overfitting. Similar to the HeiPorSPECTRAL results, MST++ excels on the MSI Brain dataset, achieving top scores across all metrics. This reinforces the efficacy of transformer-based architectures like MST++ in medical hyperspectral imaging applications.

\subsection{Model Size and Performance Comparison}
As depicted in \cref{tab:methods_comparison}, the number of parameters varies significantly among the different models (1.6K to 32.5M parameters). These differences in parameter counts across the datasets can be attributed to the higher dimensionality of the output layers. Interestingly, the largest model, UNet (32.5M parameters), does not yield the best performance in terms of metrics. Conversely, the smallest model, PixelFeatureNet (1.6K parameters), demonstrates adequate performance despite its limited parameter count. LocalFeatureNet (6.9K/9.3K) strikes a balanced trade-off between parameter size and performance. MST++ achieves the best results while maintaining significantly fewer parameters (416K/5.6M) than UNet, establishing it as the optimal choice overall.

\subsection{Reconstruction for different hyperspectral ranges}\label{sec:Reconstruction for different hyperspectral ranges}

\begin{table}[!b]
    \captionsetup{width=0.1\linewidth} % Adjust the width as needed
    \caption{Comparison of Methods on Different Datasets using\\MAE evaluated on Full, RGB, and HSI Hyperspectral ranges.\\Bold indicates best result for dataset.}
    \begin{tabularx}{\columnwidth}{>{\raggedright\arraybackslash}X >{\centering\arraybackslash}Y >{\centering\arraybackslash}Y >{\centering\arraybackslash}Y} 
        \toprule
        &  \multicolumn{3}{c}{\textbf{MSI Brain Dataset}}  \\
        \textbf{Architecture}  & \textbf{Full} & \textbf{Visible} & \textbf{Extended}  \\
        \cmidrule(lr){2-4} 
        PixelFeatureNet & 0.0225 & 0.0202 & 0.0356 \\           
        LocalFeatureNet & 0.0129 & 0.0100 & 0.0292 \\
        UNet & 0.0153 & 0.0125 & 0.0313 \\
        MST++ & \textbf{0.0121} & \textbf{0.0098} & \textbf{0.0257} \\
        \cmidrule(lr){1-4} 
        &  \multicolumn{3}{c}{\textbf{HeiPorSPECTRAL Dataset}}  \\
                \cmidrule(lr){2-4} 

         & \textbf{Full} & \textbf{Visible} & \textbf{Extended}  \\
        PixelFeatureNet & 0.0016 & 0.0014 & 0.0018 \\
        LocalFeatureNet & 0.0015 & 0.0012 & 0.0017 \\
        UNet & 0.0013 & 0.0010 & 0.0015 \\
        MST++ & \textbf{0.0012}& \textbf{0.0009} & \textbf{0.0014} \\
        \bottomrule
    \end{tabularx}
    \label{tab:ranges}
\end{table}

As previously mentioned our hypothesis was that the extended hyperspectral range poses a larger challenge. Often these higher near infrared channels are used for the extraction of functional information such as blood oxygenation or perfusion parameters \cite{Holmer2018}. Therefore the reconstruction of the extended ranges should be high to ensure that any downstream tasks are performed accurately. 

In \cref{tab:ranges}, we present a comparison of spectral reconstruction across different wavelength ranges using MAE as the evaluation metric. Our findings indicate that for the Full spectrum, results align closely with those from RMSE, except for the UNet model, which performs second best on the HeiPorSPECTRALDataset. RMSE, being more sensitive to large errors such as outliers, suggests that UNet generally exhibits larger errors despite having a lower MAE.

Furthermore, our analysis reveals consistently lower MAE values for the visible range compared to the full range, underscoring the relative ease of reconstructing wavelengths within the visible spectrum. Conversely, reconstruction in the extended range proves more challenging, with MAE values nearly three times higher on the MSI Brain Dataset when compared between visible and extended ranges.

Across both datasets, models with lower overall MAE tend to exhibit lower MAE values for visible and extended ranges as well. However, the variability in MAE between visible and extended ranges underscores the importance of analyzing errors across different spectral bands or specific wavelengths, particularly in medical applications where precise spectral information informs critical decisions in diagnosis and treatment.

\begin{figure}[h!]
    \centering
    \includegraphics[width=8cm]{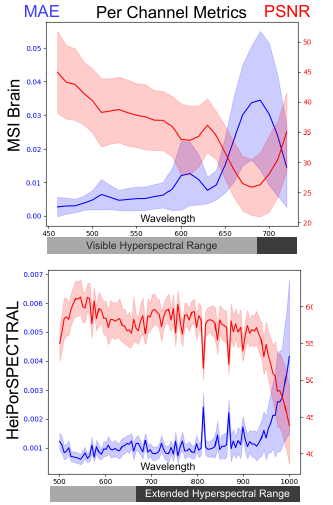}
    \captionsetup{width=\linewidth} % Adjust the width of the caption
    \caption{\textcolor{rebuttal}{Channel wise MAE and PSNR values for the best performing model MST++ on both datasets. The Mean value is represented as solid line and the standard deviation as outline.}}
 
    \label{fig:per_channel_metrics}
\end{figure}

\textcolor{rebuttal}{
For a more detailed analysis of all spectral bands, the mean and standard deviation of the MAE and PSNR metrics for both datasets are presented in \cref{fig:per_channel_metrics}. In both datasets, the reconstruction of the Extended Hyperspectral Range shows greater deviation and larger errors. There is a general decline in reconstruction performance from lower to higher wavelengths, with notable deviations in the HeiPorSPECTRAL dataset around 810 nm and in the MSI Brain dataset around 600 nm. Additionally, the MSI Brain dataset shows the worst performance around 690 nm, which can be explained by the distribution of information. As seen in \cref{fig:qualitative_MSI}, the highest values in the dataset are around 640 nm, leading to larger absolute errors. Despite some outliers, these results underscore the increasing difficulty in reconstructing signals as the wavelengths move farther from the visible Hyperspectral Range. This difficulty is particularly evident in the HeiPorSPECTRAL dataset, where there is a marked increase in mean and standard deviation around 900 nm to 1000 nm.}

\subsection{Qualitative Evaluation}
\begin{figure*}[h!]
    \centering
    \includegraphics[width=\textwidth]{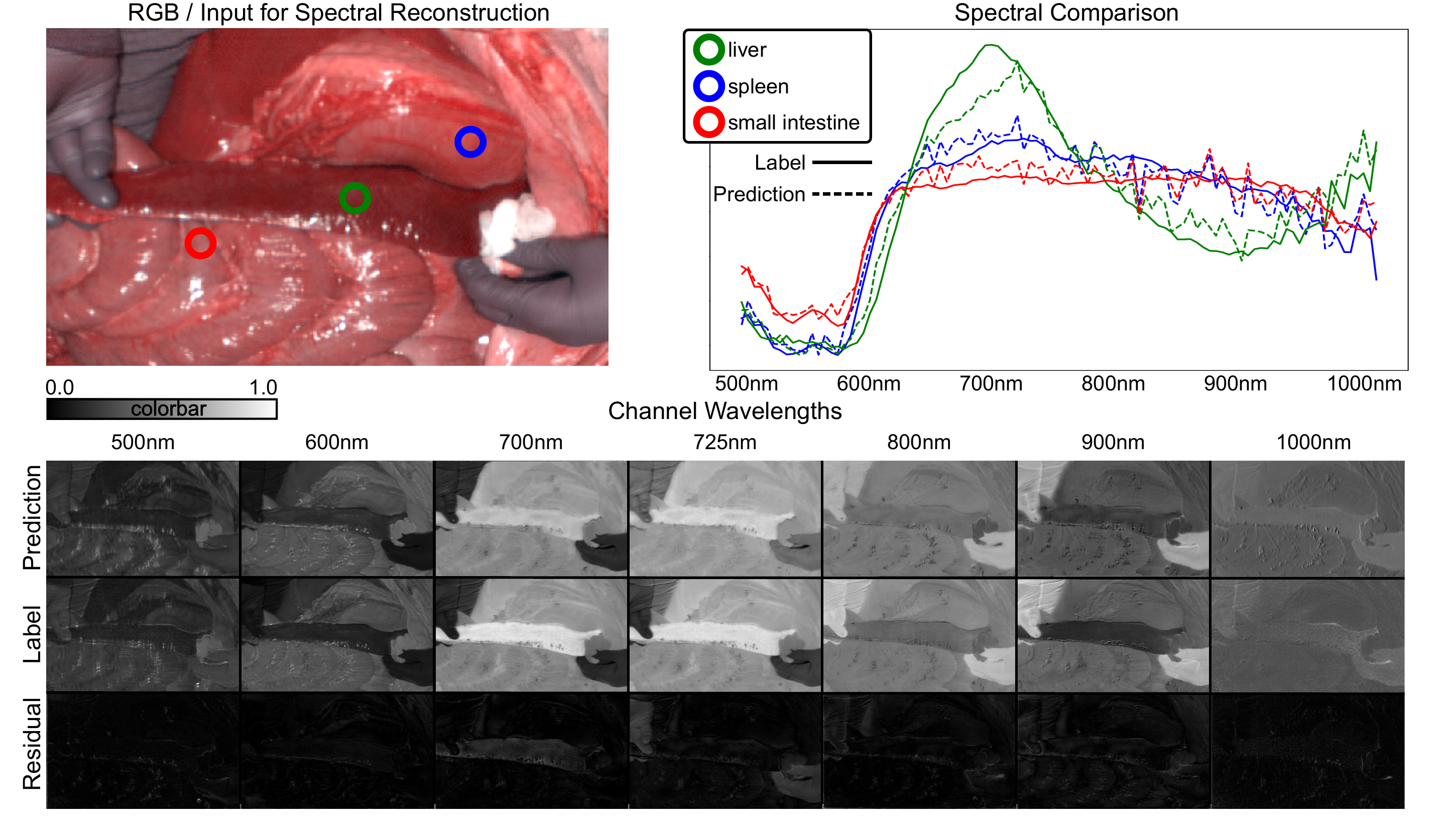}
    \captionsetup{width=\linewidth} % Adjust the width of the caption
    \caption{Qualitative Results HeiPorSPECTRAL using the best performing model MST++. The Reconstructed HSI is compared to the actual HSI values over all spectra for 3 distinct locations Gren, Blue, Red seen on the top right. At the bottom, we show the predicted HSI, label, and residual of 7 out of the 100 distinct channels to visualize the performance for each pixel.}
 
    \label{fig:qualitative_HeiPorSPECTRAL}
\end{figure*}

In \cref{fig:qualitative_HeiPorSPECTRAL}, we present a qualitative assessment of spectral reconstruction using the MST++ model on the HeiPorSPECTRAL dataset. Positioned at the top left of the figure is the RGB input image, serving as the initial input for our reconstruction pipeline. Additionally, three spectra from distinct points are displayed, each denoted by different colors to signify their spatial locations. The spectra are overlaid with dotted lines representing predicted HSI values, contrasted against solid lines indicating the ground truth labels.
Overall, the spectral curves closely track the label curves across all three sample points. The spectrum representing the Liver (green) exhibits notable discrepancies particularly around wavelengths 650 nm to 700 nm, highlighting the occasional inaccuracies inherent in spectral reconstruction, necessitating careful design.

In the spectrum of the small intestine (red), reconstruction closely mirrors the label curve, albeit with noticeable noise around 650 nm to 750 nm and more prominently near 900 nm. This observation aligns with our findings that the Extended HeiPorSPECTRAL range poses greater challenges for accurate reconstruction.

The spectrum of the spleen (blue) reveals distinguishable features between the small intestine (red) and spleen (blue) spectra, particularly evident in the 500 nm to 600 nm range and around 700 nm, with their spectra intersecting around 880 nm. This consistency between label and reconstruction suggests that reconstructed spectra could serve as valuable inputs for tissue segmentation methods, leveraging hyperspectral cameras and knowledge of tissue-specific wavelength fingerprints \cite{Studier-Fischer2022} without the need for manual labeling.

At the bottom of the figure, we additionally display seven out of the 100 distinct channels. It is observed that the residual error is less noticeable in the lower wavelengths around 500 nm and 600 nm compared to the range between 700 nm and 900 nm. Visually, the predicted quality of these channels appears convincing, although there are notable differences such as the presence of gloves at wavelength 725 and discrepancies in the liver spectrum at wavelength 700. Nevertheless, overall, the results indicate a low residual error between predicted and labeled spectra, thereby validating the findings from the quantitative metrics.

\begin{figure*}[h!]
    \centering
    \includegraphics[width=\textwidth]{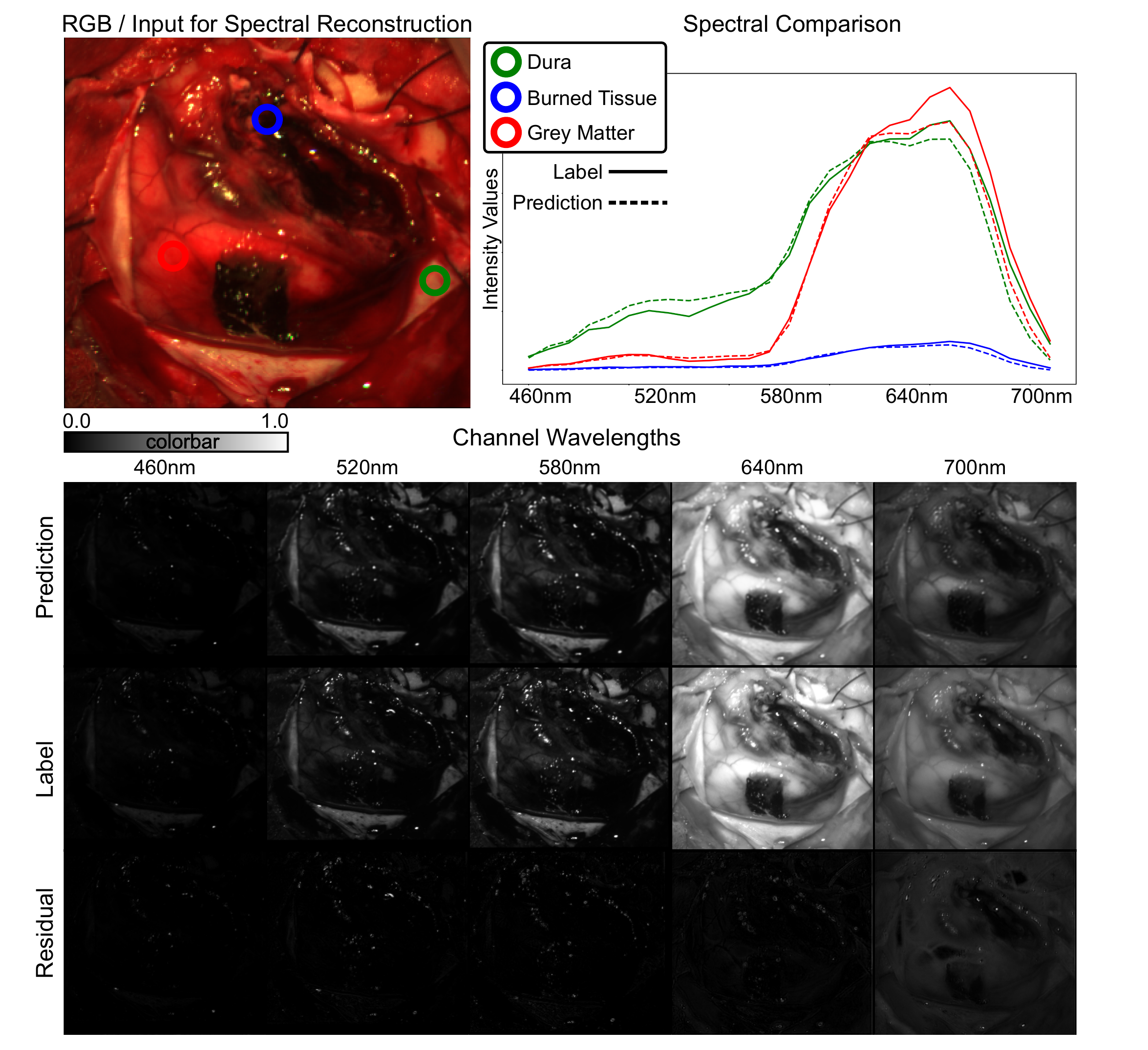}
    \captionsetup{width=\linewidth} % Adjust the width of the caption
    \caption{Qualitative Results MSI Dataset. The Reconstructed HSI is compared to the actual HSI values over all spectra for 3 distinct locations Green, Blue, and Red seen on the top right. At the bottom, we show the predicted HSI, label, and residual of 6 out of the 100 distinct channels to visualize the performance for each pixel.}
    \label{fig:qualitative_MSI}
\end{figure*}

For the MSI Brain Dataset, the predicted spectra closely resemble the labeled spectra across various tissue types as seen in \cref{fig:qualitative_MSI}. Again, three distinct points were selected in the image to compare spectral responses. The blue circle represents burned tissue post-tumor extraction, where the spectra are almost perfectly reconstructed, likely due to high absorption across all wavelengths. At the red point, which indicates grey matter in the brain, the spectra exhibit a wide range from near-zero at 460 nm to nearly one at 640 nm. Despite some errors around the peak at 640 nm, these spectral differences are largely recovered from the RGB information. The green circle represents the dura of the brain, where the spectra are also well recovered. Notably, there is almost perfect overlap between the prediction at 640nm for the red point and the label for the green point, suggesting potential confusion in distinguishing between different tissue types during reconstruction. This underscores the importance of examining multiple wavelengths for future classification tasks based on spectra. At the bottom, we again display the predicted and labeled distinct channels. We observe that the residual error at wavelength 700 nm is generally higher compared to other wavelengths. This channel was designated to the extended range for reconstruction, highlighting the greater difficulty in reconstructing the extended spectral range.

Overall, the residual errors and point spectra demonstrate promising performance of spectral reconstruction for medical imaging from RGB using the MST++ model.

\subsection{Generalisation across Datasets}
\textcolor{rebuttal}{
Pre-training on one dataset and fine-tuning on another using transfer learning can potentially enhance results and provide insights into the generalization of methods across different datasets \cite{Yosinski2014}. This capability is particularly valuable given the limited availability of datasets and the frequent use of pre-training in the literature for RGB. This process is particularly challenging due to the disparity in output channels between the HeiPorSPECTRAL dataset, which has 100 channels, and the MSI Brain dataset, which has 27 channels. To address the mismatch in output dimensions, we replaced the classification head to align with the target number of output. We fine-tuned the models for a maximum of 50 epochs choosing the best results on the validation dataset to select the best performing model. We then run the test on the orginial test set of the respective target dataset. The results are detailed in \cref{tab:transfer_learning}. 
}

\textcolor{rebuttal}{
We observed that both the PixelFeatureNet and the Local FeatureNet did not benefit from transfer learning, as evidenced by poorer performance on all four metrics. This could be related to the relatively small number of learnable weights in these architectures, where the replacement of the classification head represents a substantial proportion of the total weights. However, for the PixelFeatureNet, the SSIM metric was higher for the MSI Brain dataset when pre-trained, suggesting that transfer learning might aid in producing more realistic outputs in this specific case. The UNet architecture shows some improvements with transfer learning: SAM metrics improve on the MSI Brain dataset, while RMSE and SSIM improve on the HeiPorSPECTRAL dataset. This suggests that incorporating information from the additional dataset has been beneficial. For the MST++ model, transfer learning also leads to improvements in SAM for the HeiPorSPECTRAL dataset. The benefits are more pronounced, with notable enhancements across RMSE (\textminus0.0004), PSNR (+0.1178), and SSIM (+0.0070) for this dataset. In contrast, other architectures exhibit inconsistent results with transfer learning, showing a mix of metric improvements and deteriorations without a clear pattern of which metrics are consistently improved. These results indicate that the transformer-based MST++ architecture benefits the most from transfer learning. Moreover, transferring from the higher spectral channels of HeiPorSPECTRAL to the lower channels of MSI Brain generally seems to yield better results. However, these findings are preliminary, and more advanced approaches to transfer learning should be explored. This work represents an initial step toward understanding the effectiveness of transfer learning in this context.
}

\begin{table*}[h!]
    \centering
    \caption{\textcolor{rebuttal}{Comparison of transfer learning results across different datasets is presented using RMSE$\downarrow$, SAM$\downarrow$, SSIM$\uparrow$, and PSNR$\uparrow$ metrics. In this context, $\downarrow$ indicates that lower values are better, while $\uparrow$ signifies that higher values are preferable. Bold text highlights the superior metric values for each dataset and metric, comparing results between cases of no pre-training and pre-training on another dataset, such as from MSI Brain (source) to HeiPorSPECTRAL (target) and vice versa.}
}
    \begin{tabularx}{\textwidth}{>{\raggedright\arraybackslash}X >{\centering\arraybackslash}Y | > {\centering\arraybackslash}Y >{\centering\arraybackslash}Y | >{\centering\arraybackslash}Y >{\centering\arraybackslash}Y } 
        \toprule
        \textbf{Architecture}  & \multicolumn{1}{c}{\textbf{Metric}} & \multicolumn{2}{c}
        {\textbf{MSI Brain}} & \multicolumn{2}{c}{\textbf{HeiPorSPECTAL}} \\
        \midrule
        \multicolumn{2}{c}{\textbf{Pre-trained}} & \textit{None} & HeiPorSPECTRAL & \textit{None} & MSIBrain \\
        \midrule
        \multirow{4}{*}{PixelFeatureNet} & RMSE & \textbf{0.0415} & 0.0769 & \textbf{0.0026} &  0.0031 \\
         & SAM & \textbf{0.1985} & 0.3128& \textbf{0.1703} &  0.2318 \\
         & PSNR & \textbf{28.1059} & 26.9092 & \textbf{52.2938} &  50.3886 \\
         & SSIM & 0.7641 & \textbf{0.8144} & \textbf{0.9841} &  0.9750 \\
        \midrule
        \multirow{4}{*}{LocalFeatureNet} & RMSE & \textbf{0.0336} & 0.0386 & \textbf{0.0025} &  0.0026 \\
         & SAM & \textbf{0.1620 }& 0.1811 & \textbf{0.1578} &  0.1758 \\
         & PSNR & \textbf{30.0913} & 28.9054 & \textbf{52.710}9 &  52.0930 \\
         & SSIM & \textbf{0.9484} & 0.9079 & \textbf{0.9859} &  0.9843 \\
        \midrule
        \multirow{4}{*}{UNet} & RMSE & \textbf{0.0376} & 0.0436 & 0.0029 & \textbf{ 0.0026} \\
         & SAM & 0.1861 & \textbf{0.1811} & \textbf{0.1258} &  0.1763 \\
         & PSNR & \textbf{29.0298} & 28.0845 & \textbf{52.5472} &  52.1680 \\
         & SSIM & \textbf{0.9334} & 0.9078 & 0.9498 &  \textbf{0.9868} \\
        \midrule
        \multirow{4}{*}{MST++} & RMSE & 0.0312 & \textbf{0.0308} & \textbf{0.0021} &  \textbf{0.0021} \\
         & SAM & \textbf{0.1528} & 0.1592 & 0.1425 &  \textbf{0.1368} \\
         & PSNR & 30.5883 & \textbf{30.7061} & \textbf{53.9625} &  53.8199 \\
         & SSIM & 0.9507 & \textbf{0.9573} & \textbf{0.9924} &  0.9920 \\
         \bottomrule
    \end{tabularx}
    \label{tab:transfer_learning}
\end{table*}

\section{Conclusion}
In this study, we have explored various approaches for spectral reconstruction from RGB to hyperspectral images. Our findings underscore several key insights into the effectiveness of different model architectures and strategies for improving reconstruction accuracy.

Firstly, we demonstrated that integrating spatial information significantly enhances reconstruction quality. Models like MST++ and LocalFeatureNet, which incorporate spatial context into their spectral predictions, consistently outperformed simpler models like PixelFeatureNet. This highlights the importance of leveraging spatial relationships to refine spectral predictions across diverse tissue types and imaging conditions. More parameters not necessarily translates to better reconstruciton as evident from UNet. However, the modeling power of transformers used in MST++ translates to best results validated on different surgical settings on two dataset.

\textcolor{rebuttal}{
Furthermore, while our results demonstrate strong performance in reconstructing visible wavelength ranges, we recognize the challenges associated with extending this reconstruction to broader spectral ranges. Absorption spectra typically extend across a wide spectral region \cite{Cheong1990}, therefore it is possible, in principle, to infer information from the IR or UV ranges from responses in the visible. However, the difficulties encountered in accurately capturing spectral information at longer wavelengths highlight the need for further investigation and refinement, especially for extracting functional insights critical to medical diagnostics.}

Moreover, our investigation revealed that spectral reconstruction offers a promising label-free approach for tissue classification in medical imaging. By reconstructing hyperspectral signatures from RGB data, we can exploit spectral fingerprints to classify tissues accurately. This capability not only simplifies data acquisition but also opens avenues for automating diagnostic processes and improving treatment decisions.

\textcolor{rebuttal}{Our results suggest that transfer learning has the potential to enhance spectral reconstruction by pre-training on one spectral dataset and then fine-tuning on a target dataset. We observe varying degrees of effectiveness across different architecture types, with transformers demonstrating the most significant improvements.}

In conclusion, our study underscores the transformative potential of spectral reconstruction in medical imaging, driven by advancements in spatially informed models and insights into leveraging RGB data for hyperspectral analysis. Future research directions should focus on refining reconstruction techniques to extract comprehensive functional information and further validating these approaches across broader medical applications.

\section{Ethical Approval}
\textcolor{rebuttal}{To acquire the MSI Brain dataset, the application received approval from the South Central - Berkshire B Research Ethics Committee, on behalf of the HRA, on the 04/02/2020, REC reference 19/SC/0583, IRAS ID 258210}.

\section{Acknowledgements}
\textcolor{rebuttal}{
Dr Maria Leiloglou was financially supported by the Royal Academy of Engineering (Enterprise Fellowship) and the Greek Foundation for Education and European Culture (IPEP). Mr Giulio Anichini (George Pickard’s Research Fellowship) and Mr Kevin O’Neill were financially supported by Brain Tumour Research (BTR) and  Brain Tumour Research Campaign (BTRC).}

% \begin{table*}[h!]
%     \centering
%     \caption{Comparison of Methods on Different Datasets using Root Square Error (RMSE)$\downarrow$, Spectral Angle Mapping (SAM)$\downarrow$, Structure Similarity Measure (SSIM)$\uparrow$, and Peak Signal-to-Noise Ratio (PSNR)$\uparrow$. $\downarrow$ indicates that lower values are better and $\uparrow$ indicates that higher values are better. Bold indicates best result for dataset. Additionally the number of learnable parameters is shown for neurosurgical Dataset}
%     \begin{tabularx}{\textwidth}{>{\raggedright\arraybackslash}X >{\centering\arraybackslash}Y >{\centering\arraybackslash}Y >{\centering\arraybackslash}Y } 
%         \toprule
%         \textbf{Architecture} & \textbf{\# Parameters} &\textbf{Metric} & \textbf{MSI Brain}  \\
%         \midrule

%         \midrule
%         \multirow{4}{*}{EnSurge} & \multirow{4}{*}{416K} & RMSE & \textbf{0.0312}  \\
%         & & SAM & \textbf{0.1528} \\
%         & & PSNR & \textbf{30.5883}   \\
%         & & SSIM & \textbf{0.9507}  \\
%         \bottomrule
%     \end{tabularx}
%     \label{tab:methods_comparison}
% \end{table*}


\bibliography{bibtex}  % Specify the BibTeX file
\bibliographystyle{plain}
\end{document}